\documentstyle[prd,aps,twocolumn,epsf]{revtex}

\def\IN{\relax{\rm I\kern-.18em N}}
\def\IR{\relax{\rm I\kern-.18em R}}

\font\cmss=cmss12 \font\cmsss=cmss12 at 7pt
\def\IZ{\relax\ifmmode\mathchoice
{\hbox{\cmss Z\kern-.4em Z}}{\hbox{\cmss Z\kern-.4em Z}}
{\lower.9pt\hbox{\cmsss Z\kern-.4em Z}}
{\lower1.2pt\hbox{\cmsss Z\kern-.4em Z}}\else{\cmss Z\kern-.4em Z}\fi}

\def\inbar{\,\vrule height1.5ex width.4pt depth0pt}
\def\IC{\relax\hbox{$\inbar\kern-.3em{\rm C}$}}

\preprint{IC preprint number}

\begin{document}

\typeout{--- Title page start ---}

\renewcommand{\thefootnote}{\fnsymbol{footnote}}


\title{Phase Transitions in Condensed Matter and Relativistic QFT
\thanks{Talk presented at 5th International
Workshop on Thermal Field Theories and Their Applications,
Regensberg, August 1998}}

\author{R.\ J.\ Rivers}

\address{Blackett Laboratory, Imperial College, London SW7 2BZ}


\maketitle

\renewcommand{\thefootnote}{\arabic{footnote}}

\setcounter{footnote}{0}

\typeout{--- Main Text Start ---}

\begin{abstract}

Kibble and Zurek have provided a unifying picture for the onset of
phase transitions in relativistic QFT and condensed matter systems respectively,
strongly supported by agreement with condensed matter experiments in
$^{3}He$. The failure of a recent experiment on $^{4}He$ to agree
with the predictions has prompted a reappraisal of this picture.  We
provide an alternative picture that explains the experimental evidence.

\end{abstract}


\narrowtext

\section{The onset of phase transitions}

The two environments in which
thermal field theory is relevant are the early universe and 
heavy-ion collisions, in both of which there are changes of phase. Not
surprisingly, one of 
the themes of this series of Workshops on Thermal Field
Theory (TFT) has been phase transitions.  

Consider a situation in which the  symmetry group  of the theory is
broken, on cooling through the critical temperature $T_c$, 
 by its degenerate groundstate manifold.
Using the tools of {\it equilibrium} TFT we can determine
the nature of the transition.
At late times after the transition the fields are ordered on large
scales, in that the field will adopt a single value from this
degenerate set over a large spatial region.  This might seem to be
all that we need to know, but  
how this is achieved can have interesting consequences. 

Such problems require that we go beyond equilibrium TFT.
In practice, we often know remarkably little about the dynamics of
thermal systems.
For
simplicity, I shall assume scalar field  order parameters, with {\it
continuous} transitions.
In particular,
I want to discuss some aspects of the {\it onset} of
such phase transitions, the very early
times after the implementation of  a  transition when
the scalar
fields are only just beginning to become ordered.  

A simple question to ask is the following: In principle, the field
correlation length diverges at a continuous transition.  In
practice, it does not.  What happens?  This
is relevant for transitions that
leave topological defects like walls, monopoles, vortices, or textures in
their wake since we might expect 'defects' to be just that, entities
whose separation is characterised by the correlation length.  If this
were simply so, a measurement of defect densities would be a
measurement of correlation lengths. 
Estimates of this early field ordering
have been made by
Kibble\cite{kibble1,kibble2}, using simple causal arguments,
 because of the implication of defects for astrophysics.
Vortices, in
particular (cosmic strings), can be important for structure formation
in the universe. 
\footnote{In a different context, the topological textures of the NLSM $O(4)$ theory can be
identified with Skyrmionic hadrons produced in the hadronisation of
quark-gluon plasma.}  

There are great 
difficulties in converting predictions for the early universe into
experimental observations.
Zurek suggested\cite{zurek1} that similar arguments were applicable
to condensed matter systems for which direct experiments on defect
densities could be
performed.  This has lead to a burst of activity from theorists
working on the boundary between QFT and Condensed Matter theory and
from condensed matter experimentalists. To date almost all experiments have involved
superfluids, with scalar order parameters of the type considered
above and all but one experiment is in agreement with these simple
causal predictions.

In this talk I shall
\begin{itemize}
\item
review the Kibble/Zurek causality predictions for initial correlation lengths and defect
densities and summarise the results of the condensed matter
experiments.
\item
present an alternative picture for the onset of defect production,
applicable to both relativistic quantum fields and condensed matter systems.
\item
show how this alternative picture gives essentially the same results
as the Zurek picture for those condensed matter systems for which
there is experimental agreement.
\item
provide an explanation for why some condensed matter experiments
will be in disagreement with Zurek's predictions, including the
experiment in question.
\item
try to draw some broader conclusions about the relevance of Kibble's
predictions for phase transitions in QFT.
\end{itemize}

\section{When symmetry breaks, how big are the smallest identifiable pieces?}

Insofar that defect density can be correlated simply to correlation
length, the {\it maximum} density (an experimental observable in
condensed matter systems, although not for the early universe) 
will be associated with the {\it smallest} identifiable
correlation length in the broken phase once the transition has been effected.
This provides the initial condition for the evolution of field ordering.
In order to see how to  identify these
'smallest pieces'\footnote{The
title of this section is essentially that posed in recent papers by
Zurek\cite{zurek2}.} it is sufficient to consider the simplest theory, that of a
single relativistic quantum scalar field in three spatial dimensions
with a double well potential, undergoing a
temperature quench.  In the first
instance we assume  
that the qualitative dynamics
of the transition are conditioned by the field's
{\it equilibrium} free energy, of the form
\begin{equation}
F(T) = \int d^{3}x\,\,\bigg(\frac{1}{2}(\nabla\phi)^{2}
+ \frac{1}{2}m^{2}(T)\phi^{2} + \frac{1}{4}\lambda \phi^{4}\bigg).
\label{FR}
\end{equation}
Prior to the transition, at temperature $T>T_{c}$, the critical
temperature, $m (T) > 0$ plays the role of an effective 'plasma'
mass due to the interactions of $\phi$ with the heatbath, which
includes its own particles. 
After the transition, when $T$ is effectively zero, $m^{2}(0) = - M^{2}< 0$
enforces the $Z_{2}$ symmetry-breaking, with field expectation
values $\langle\phi\rangle = \pm\eta$, $\eta^{2} = M^{2}/\lambda$.  
The change in temperature that leads to the change in the sign of
$m^{2}$ is most simply understood as a consequence of the system expanding.
Thus, in the early universe, a weakly interacting relativistic
plasma at temperature $T\gg M$ has an entropy density $s\propto T^{3}$.
As long as thermal equilibrium can be maintained, constant 
entropy $S$ per comoving volume,  $S\propto s a(t)^{3}$, gives
$T\propto a(t)^{-1}$ and falling, for increasing scale factor $a(t)$.
 We shall be too simple in our discussion to warrant
the inclusion of a metric in Eq.\ref{FR}.
  
Some caution is needed in
interpreting $V(\phi) =m^{2}(T)\phi^{2}/2 + \lambda\phi^{4}/4$ 
as the conventional Effective Potential
since this is a zero-momentum, or infinite-volume field average,
construct.
At the very least, a more realistic potential
for dynamics would be coarse-grained to the relevant scale.  In
practice, as long as we never look on scales smaller that the
field correlation length this turns out not to matter much.  
The only relevant point is that
$m^{2}(T)$ vanishes at $T=T_{c}$, true however $V(\phi )$ is defined.  

The purpose of $V(\phi , T)$ is, initially, to set the scale of the
fluctuations of $\phi$ about $\phi = 0$, the subsequent false vacuum.
The {\it equilibrium}
correlation length of the field fluctuations is $\xi_{eq}(T)= |m^{-1}
(T)|$. It is sufficient to
adopt a mean-field approximation in which $m^{2}(T)\propto (T-T_{c})$. 
 For fields averaged on the scale of
$\xi_{eq}(T)$ we find
$\langle \phi^{2}\rangle = O(m(T)T)$, much less than $\eta^{2}$ for
our relativistic theory in which $T_{c} = O(\eta )$.  
After the transition the
field, at each point in space, begins to 
collapse to one of the false vacua $\phi = \pm\eta$,
$\eta^{2} = M^{2}/\lambda$, as signalled by $V(\phi )$, 
initially in a random way.  However, the field soon begins to become
organised into 'domains', in each of which the field has a constant sign.
The defects of this simple model are domain walls, across which the
field flips from one minimum to the other.
Defect density then looks to be related to domain size.

How this collapse takes place determines the size of the first
identifiable domains.  It was suggested by Kibble and Zurek 
that this size is
essentially the field correlation length $\xi_{eq}$ at some appropriate
temperature close to the transition.  
I shall argue later that this is too simple but, nonetheless
it is a plausible starting point.  Near the continuous transition $\xi_{eq}$ becomes
large, approaching infinity at $T_{c}$.  In practice physical
correlations remain finite.  The question then becomes: how large
does the field correlation get in practice?  Two very different 
mechanisms have been
proposed for estimating this size.

\subsection{Thermal activation}

In the early work on the cosmic string scenario it was
assumed\cite{kibble1}  that
initial domain size was fixed in the Ginzburg regime.  By this we mean the
following.  Suppose the temperature $T(t)$ varies sufficiently slowly
with time $t$ that it makes sense to replace $V(\phi ,T)$ by $V(\phi
, T(t))$.  Well away from the transition this is justified, but
close to the transition it is not.  A band in temperature about
$T_{c}$   in which it is not sensible is $T^{-}_{G} < T < T^{+}_{G}$,
where $T^{\pm}_{G}$ are the Ginzburg temperatures.  As we cool down
in the vicinity of $T_{c}$, $V(\phi ,T)$ ceases to be a reliable guide to the
field fluctuations at $T^{+}_{G}$, where the relative Ginzburg temperature
is of order of
magnitude $|1- T^{+}_{G}/T_{c}| = O(\lambda )$. 
Once we are below $T_{c}$, and the central hump in $V(\phi
, T(t))$ is forming,
$T^{-}_{G}$ signals the temperature above which there is a
significant probability for thermal fluctuations over the central
hump on the scale of the correlation length.  Most simply, it is
determined by the condition
\begin{equation}
\Delta V(T^{-}_{G})\xi_{eq}^{3}(T^{-}_{G})\approx T^{-}_{G}
\end{equation}  
where $\Delta V(T)$ is the difference between the central maximum
and the minima of $V(\phi ,T)$.  Again we find 
\newline
$|1-T^{-}_{G}/T_{c}| = O(\lambda )$.

Whereas, above $(T^{-}_{G})$ there will be a population of
'domains', fluctuating in and out of existence,
at temperatures below $ T^{-}_{G}$ fluctuations from one minimum to
the other become increasingly unlikely. When this happens the
correlation length is 
\begin{equation}
\xi_{eq}(T^{-}_{G}) = 
O\bigg(\frac{\xi_{0}}{\sqrt{\lambda}}\bigg),
\label{xiG}
\end{equation}
where $\xi_{0} = M^{-1}$ is the natural unit of length, the Compton
wavelength of the $\phi$ particles. 

It is tempting to identify $\xi_{eq}(T^{-}_{G})$
with the scale at which stable domains begin to form. Subsequent
work shows this to be incorrect.  Thermal
fluctuations {\it are} relevant to the formation of small domains, and to
wiggles in the boundaries of larger domains, but not in the
formation of larger domains themselves.  That is an issue that
requires more than equilibrium physics.  The most simple dynamical arguments
can be understood in terms of causality.

\subsection{Causality}

If the Ginzburg criteria attempt to set scales once the critical
temperature has been {\it passed}, causal arguments attempt to set scales
{\it before} it is reached. 
We have seen that $\xi_{eq}(T(t))$ diverges at $T(t) =T_{c}$, which
we suppose  happens at $t=0$. This
cannot be the case for the true correlation length $\xi (t)$, which can
only grow so far in a finite time.  Initially, for $t<0$, when we are far from
the transition, we again assume effective equilibrium, and the field
correlation length $\xi (t)$ tracks $\xi_{eq}(T(t))$ approximately. 
However, as we get closer to the transition $\xi_{eq}(T(t))$ begins
to increase arbitrarily fast.  As a crude upper bound, the true
correlation length fails to keep up with $\xi_{eq}(T(t))$
by the time $-t_{C}$ at which $\xi_{eq}$ is growing at the speed of light,
$d\xi_{eq}(T(-t_{C}))/dt =1$.  It was suggested by Kibble\cite{kibble2} that,
once we have reached this time $\xi (t)$ {\it freezes} in, remaining approximately
constant until the time $t\approx +t_{C}$ after the transition
when is once again becomes comparable to the now {\it decreasing}
value of $\xi_{eq}$. The correlation length $\xi_{eq}(t_{C})=\xi_{eq}(-t_{C})$ is
argued to provide the scale for the minimum domain size {\it after} the
transition.

Specifically, if we assume a time-dependence $m^{2}(t) =
-M^{2}t/t_Q$ in the vicinity of $t=0$, when the transition begins to
be effected, then the causality condition gives
$t_C=t_{Q}^{1/3}(2M)^{-2/3}$.  As a result,
\begin{equation}
M\xi_{eq}(t_{C}) = (M\tau_{0})^{1/3},
\label{xiC0}
\end{equation}
which, with condensed matter in mind, we write as
\begin{equation}
\xi_{eq}(t_{C}) = \xi_{0}\bigg(\frac{\tau_{Q}}{\tau_{0}}\bigg)^{1/3}
\label{xiC}
\end{equation}
where $\tau_{0}  = \xi_{0}= M^{-1}$ are the natural time and
distance scales.  In
contrast to Eq.\ref{xiG}, Eq.\ref{xiC} depends explicitly on the
quench rate, as we would expect.  For $\tau_{Q}\gg \tau_{0}$ the
field is correlated on a scale of many Compton wavelengths.

\subsection{QFT or Condensed Matter}

This approach of Kibble was one of the motivations for a similar
analysis by Zurek\cite{zurek1} of transitions with scalar order parameters in condensed matter.
Qualitatively, neither the Ginzburg thermal fluctuations, with
fluctuation length Eq.\ref{xiG}, nor the simple causal
argument above depend critically on the fact that the free energy Eq.\ref{FR}
is originally assumed to be derived from a relativistic
{\it quantum} field theory.  After rescaling, $F$ could equally well be the Ginzburg-Landau
free energy 
\begin{equation}
F(T) = \int d^{3}x\,\,\bigg(\frac{\hbar^{2}}{2m}(\nabla\phi )^{2}
+\alpha (T)\phi^{2} + \frac{1}{4}\beta \phi^{4}\bigg)
\label{FNR}
\end{equation}
for a non-relativistic condensed matter field,
in which $\alpha (T)\propto m^{2}(T)$ vanishes at
the critical temperature $T_{c}$.  The only difference is that, in
the causal argument, the speed of light should be replaced by the
speed of sound,
with different critical index.

Explicitly, let us assume the mean-field result 
$\alpha (T) = \alpha_{0}\epsilon (T_{c})$, where $\epsilon = (T/T_{c}
-1)$, remains valid as $T/T_{c}$ varies with time $t$.
In particular, we first take  $\alpha (t)=\alpha (T(t))=-\alpha_{0}
t/\tau_{Q}$ in the vicinity of $T_{c}$.  Then the fundamental length and
time scales $\xi_{0}$ and $t_0$ are given from Eq.\ref{FNR} as
 $\xi_{0}^{2} = \hbar^{2}/2m\alpha_{0}$ and $\tau_{0} = \hbar
/\alpha_{0}$. It follows that 
the
equilibrium correlation length $\xi_{eq} (t)$ and the relaxation
time $\tau (t)$ diverge when $t$ vanishes as 
\begin{equation}
\xi_{eq} (t) = \xi_{0}\bigg|\frac{t}{\tau_{Q}}\bigg|^{-1/2},
\,\,\tau (t) = \tau_{0}\bigg|\frac{t}{\tau_{Q}}\bigg|^{-1}.
\end{equation}
The speed of sound is $c(t) =\xi_{eq} (t)/\tau (t)$, slowing down as
we approach the transition as $|t|^{1/2}$.  The causal counterpart
to $d\xi_{eq}
(t)/dt = 1$ for the relativistic field is
$d\xi_{eq}
(t)/dt = c(t)$.  This is satisfied at $t=-t_{C}$, where $t_{C}
=\sqrt{\tau_{Q}\tau_{0}}$, with corresponding correlation length
\begin{equation}
\xi_{eq}(t_{C}) =\xi_{eq}(-t_{C}) = \xi_{0}\bigg(\frac{\tau_{Q}}{\tau_{0}}\bigg)^{1/4}.
\label{xiZ}
\end{equation}
(cf. Eq.\ref{xiC}).
A variant of this argument that gives essentially the same results
is obtained by comparing the quench rate directly to the relaxation rate of
the field fluctuations.  We stress that, yet again, the assumption is that the
length scale that determines the initial correlation length of the
field freezes in {\it before} the transition begins.
Whatever, the field is already correlated on a scale of many Compton
wavelengths when it begins to unfreeze.

\section{Experiments}

The end result of the simple causality arguments is that, both for
QFT and condensed matter, when the
field begins to order itself its correlation length has the form
\begin{equation}
\xi_{eq}(t_{C}) = \xi_{0}\bigg(\frac{\tau_{Q}}{\tau_{0}}\bigg)^{\gamma}.
\label{xiKZ}
\end{equation}
for appropriate $\gamma$.  \footnote{In fact, the powers of
Eq.\ref{xiC} and Eq.\ref{xiZ} are mean-field results, changed on
implementing the renormalisation group.}  

For a relativistic {\it complex} scalar field $\phi$ with a global $U(1)$ or $O(2)$
symmetry the causality argument is the same, but the physical
situation is very different.  Consider the theory 
controlled
by a free energy 
\begin{equation}
F(T) = \int d^{3}x\,\,\bigg(|\nabla\phi|^{2}
+ m^{2}(T)|\phi|^{2} + \lambda |\phi|^{4}\bigg)
\label{FR2}
\end{equation}
with $m^{2}(T)$ switching from  positive to negative.
The minima of the final  potential of
Eq.\ref{FR2} now constitute the circle $\phi = \eta e^{i\alpha}$.  
When the transition begins $\phi$ begins to fall into the valley of
the potential, choosing a random phase.  This randomly chosen phase
will vary from point to point. 
Solutions to $\delta F/\delta\phi = 0$ include vortices,
topological defects,
tubes of 'false' vacuum $\phi\approx 0$, around which the field
phase changes by $\pm 2\pi$.     
In an early universe context these are
'cosmic strings', but we shall not consider their properties here.
 The jump that Kibble made was to assume that the correlation
 length Eq.\ref{xiC}, equally applicable to a complex field,
 also sets the scale for the typical minimum intervortex
distance.  

That is, the {\it initial} vortex density
$n_{def}$ is
\begin{equation}
n_{def} = O\bigg(\frac{1}{(\xi_{eq}(t_{C}))^{2}}\bigg)= 
O\bigg(\frac{1}{\xi_{0}^{2}}\bigg(\frac{\tau_{0}}{\tau_{Q}}\bigg)^{2\gamma}\bigg).
\label{ndef}
\end{equation}
for $\gamma = 1/3$.
Equivalently, the length of vortices in a box volume $v$ is $O(n_{def}v)$.
We stress that this assumption is {\it independent} of the argument
that lead to Eq.\ref{xiC}.

Since $\xi_{0}$ also measures cold vortex thickness, $\tau_{Q}\gg
\tau_{0}$ corresponds to a measurably large number of widely
separated vortices.
If it could be argued that this initial network behaves classically
then, thereafter, the density will reduce due to the collapse of small
loops, intersections chopping off loops which in turn collapse, and vortex straightening so as to
reduce the gradient energy of the field.

Even if cosmic strings were produced in so simple a way in the very
early universe it is not possible to compare the density Eq.\ref{ndef} with experiment. 
  It was Zurek who
first suggested that this causal argument for defect density be
tested in condensed matter systems, particularly in liquid helium.

\subsection{Superfluid helium}

Vortex lines in both superfluid $^{4}He$ and $^{3}He$ are good analogues of global
cosmic strings. A crude but effective model is to treat the system
as composed of two fluids, the normal fluid and the superfluid,
which has zero viscosity. In $^{4}He$ the bose superfluid is
characterised by a complex field $\phi$, whose squared modulus
$|\phi |^{2}$ is the superfluid density.  The superfluid fraction is
unity at absolute zero, falling to zero as the temperature rises to
the lambda point at 2.17K.  The Landau-Ginzburg theory for $^{4}He$
has, as its free energy,
\begin{equation}
F(T) = \int d^{3}x\,\,\bigg(\frac{\hbar^{2}}{2m}|\nabla\phi |^{2}
+\alpha (T) |\phi |^{2} + \frac{1}{4}\beta |\phi |^{4}\bigg),
\label{FNR2}
\end{equation}
the scaled counterpart of Eq.\ref{FR2}.
The static classical field equation $\delta F/\delta\phi = 0$ has
vortex solutions as before, with width $\xi_{0}$.

The situation is more complicated, but more interesting, for
$^{3}He$, which becomes superfluid at the much lower
temperature of 2 mK. The reason is that the $^{3}He$ is a {\it
fermion}.  Thus the mechanism for superfluidity is very different
from that of $^{4}He$.  Somewhat as in a BCS superconductor, these
fermions form the counterpart to Cooper pairs.  However, whereas the
(electron) Cooper pairs in a superconductor form a $^{1}S$ state,
the $^{3}He$ pairs form a $^{3}P$ state. The order parameter
$A_{\alpha i}$ is a complex $3\times 3$ matrix $A_{\alpha i}$.
There are two distinct superfluid phases, depending on how the
$SO(3)\times SO(3)\times U(1)$ symmetry is broken. If
the normal fluid is cooled at low pressures, it makes a transition
to the $^{3}He-B$ phase, in which $A_{\alpha i}$ takes the form
$A_{\alpha i} = R_{\alpha i}(\omega)e^{i\Phi}$, where $R$ is a real rotation
matrix, corresponding to a rotation through an arbitrary $\omega$\cite{volovik2}

The Landau-Ginzburg free energy is, necessarily, more complicated,
but the effective potential $V(A_{\alpha i}, T)$ has the diagonal
form $V(A, T) = \alpha
(T)|A_{ai}|^{2} + O(A^{4})$ for small fluctuations, and this is all
that we need for the production of vortices at very early times.  Thus the Zurek analysis leads to
the prediction Eq.\ref{ndef}, as before, for appropriate $\gamma$.  
However, for $^{3}He$ the
mean-field approximation is good and the mean-field critical index
$\gamma = 1/4$ is not
renormalised, whereas for $^{4}He$ a better value is $\gamma = 1/3$.

\subsection{Experiments in $^{3}He$.}

Although $^{3}He$ is more complicated to work with, the experiments to check
Eq.\ref{ndef} are cleaner for both experimental and theoretical
reasons. First, because the nucleus has spin $1/2$, even individual vortices can be detected by
magnetic resonance. Second, because vortex width is many atomic
spacings the Landau-Ginzburg theory is good.

So far, experiments have been of two types.  In the Helsinki
experiment\cite{helsinki} superfluid $^{3}He$ in a rotating cryostat is bombarded by slow
neutrons.  Each neutron entering the chamber releases 760 keV, via
the reaction $n + ^{3}He\rightarrow p + ^{3}He + 760 keV$.  The
energy goes into the kinetic energy of the proton and triton, and is
dissipated by ionisation, heating a region of the sample above its
transition temperature.  The heated region then cools back through
the transition temperature, creating vortices. Vortices above a
critical size (dependent on the angular velocity of the cryostat)
grow and migrate to the centre of the apparatus, where they are counted by
an NMR absorption measurement.  Suffice to say that the quench is
very fast, with $\tau_{Q}/\tau_{0} = O(10^{3})$.   Agreement with
Eq.\ref{ndef} and Eq.\ref{xiZ} is very good, at the level of an order of magnitude.

The second type of experiment has been performed at 
Grenoble and Lancaster\cite{grenoble}.  Rather than count
individual vortices, the experiment detects the total energy going
into vortex formation. As before, $^{3}He$ is irradiated by
neutrons.  After each absorption the energy released in the form of
quasiparticles is measured, and found to be less than the total 760 keV.
This missing energy is assumed to have been expended on vortex
production.  Again, agreement with Zurek's prediction Eq.\ref{ndef} and
Eq.\ref{xiZ} is good.

\subsection{Experiments in $^{4}He$.}

The experiments in $^{4}He$, conducted at Lancaster, follow Zurek's
original suggestion.  The idea is to expand a sample of normal fluid helium, in
a container with bellows, so that it becomes superfluid at
essentially constant temperature. That is, we change $1-T/T_{c}$
from negative to positive by reducing the pressure, thereby increasing $T_{c}$. 
As the system goes into the superfluid phase a tangle of vortices is
formed, because of the random distribution of field phases.  The
vortices are detected by measuring the attenuation of second sound
within the bellows.  Second sound scatters off vortices, and its
attenuation gives a good measure of vortex density.  A mechanical
quench is slow, with $\tau_{Q}$ some tens of milliseconds, and $\tau_{Q}/\tau_{0} = O(10^{10})$.  Two
experiments have been performed\cite{lancaster,lancaster2}.  In the first fair agreement was
found with the prediction Eq.\ref{ndef}, although it was not
possible to vary $\tau_{Q}$.  However, there were potential problems
with hydrodynamic effects at the bellows, and at the capillary with
which the bellows were filled.  A second experiment, designed to
minimise these 
and other  problems has failed to see any vortices whatsoever.

There is certainly no agreement, in this or any other experiment on $^{3}He$,
with the thermal fluctuation density that would be based on
Eq.\ref{xiG}.

\section{The Kibble-Zurek picture for the freezing in of $\xi$ is correct.}

We are in the strange position that, if the second Lancaster
experiment\cite{lancaster2} is correct, either both the Zurek predictions Eq.\ref{ndef}
and Eq.\ref{xiZ}
works well, or one or both fail.  We shall now argue that the freezing
in of $\xi$ as in Eq.\ref{xiZ} is correct, essentially on
dimensional grounds.  
To show this we need a concrete model for the dynamics.  

\subsection{Condensed matter: the TDLG equation}

We assume that the dynamics
of the transition can be derived from the explicitly time-dependent
Landau-Ginzburg free energy 
\begin{equation}
F(t) = \int d^{3}x\,\,\bigg(\frac{\hbar^{2}}{2m}|\nabla\phi |^{2}
+\alpha (t)|\phi |^{2} + \frac{1}{4}\beta |\phi |^{4}\bigg).
\label{F}
\end{equation}
In (\ref{F}) $\phi = (\phi_{1} + i\phi_{2})/\sqrt{2}$ is the
complex order-parameter field, whose magnitude determines the
superfluid density.   In equilibrium at temperature
$T$, in a mean field approximation,  the chemical potential $\alpha (T)$ takes the form
$\alpha (T) = \alpha_{0}\epsilon (T_{c})$, where $\epsilon = (T/T_{c}
-1)$. 
In a quench in which $T_{c}$ or $T$ changes it is convenient to shift the origin in
time, to write $\epsilon$  as
\begin{equation}
\epsilon (t) = \epsilon_{0} - \frac{t}{\tau_{Q}}\theta (t)
\label{eps}
\end{equation}
for $-\infty < t < \tau_{Q}(1 + \epsilon_{0})$, after which
$\epsilon (t) = -1$.  $\epsilon_{0}$  
measures the original relative temperature and $\tau_{Q}$
defines the quench rate.  The quench begins at time $t = 0$ and the
transition from the normal to the superfluid phase begins at time $t
= \epsilon_{0}\tau_{Q}$.

Motivated by Zurek's later numerical\cite{zurek2} simulations with the time-dependent
Landau-Ginzburg (TDLG) equation for $F$ of (\ref{F}), 
we assume a linear response
\begin{equation}
\frac{1}{\Gamma}\frac{\partial\phi_{a}}{\partial t} = -\frac{\delta
F}{\delta\phi_{a}} + \eta_{a},
\label{tdlg}
\end{equation}
where $\eta_{a}$ is Gaussian noise. More details are given
elsewhere\cite{ray}. We can show self-consistently  that, for
the relevant time-interval $-t_{C}\leq \Delta t\leq t_{C}$
the self-interaction term can be neglected ($\beta =0$).  This both
preserves Gaussian field fluctuations and leads to 
$\xi\approx\xi_{0}(\tau_{Q}/\tau_{0})^{1/4}$ arising
in a natural way, as we shall see.  

At relative time
$\Delta t$  the diagonal {\it equal-time} correlation
function $G(r,\Delta t)$ is defined by
\begin{equation}
\langle \phi_{a}({\bf r})\phi_{b}({\bf 0})\rangle_{\Delta t} =
\delta_{ab}G(r,\Delta t).
\label{diag}
\end{equation}

Anticipating that correlation lengths are
approximately frozen
in during this period it is sufficient to perform our calculations at  $\Delta t =
0$.  Eq.\ref{tdlg} is solvable, in time and space units $\tau_{0}$ and
$\xi_{0}$, to give
the Fourier transform of $G(r,\Delta t)$ as
\begin{equation}
G(k)=e^{\tau_{Q}k^{4}}\int_{\tau_{Q}k^{2}}^{\infty}dt\,e^{-t^{2}/\tau_{Q}},
\end{equation}
independent of $\epsilon_{0}$,
provided $\epsilon_{0}\tau_{Q},\,\,\epsilon_{0}^{2}\tau_{Q}\gg 1$. 
For $^{4}He$, where this is more problematical, this is valid, 
since although $\epsilon_{0}\sim 10^{-2} -
10^{-3}$ is very small, $\tau_{Q}\sim 10^{10}$ is so large. 

On dimensional grounds the correlation length of the field is
$O(\xi_{eq}(t_{C})) = O(\xi_{0}(\tau_{Q}/\tau_{0})^{1/4})$.
However, rather than have  an {\it asymptotic} fall-off of
the form $e^{-r/\xi_{eq}(t_{C}) }$ for large $r$, $G(r)\propto
\exp(-O((r/\xi_{eq}(t_{C}) )^{4/3}))$.  
Nonetheless, numerically, it is remarkably well
represented by $e^{-r/\xi_{eq}(t_{C}) }$, with coefficient {\it unity} in the exponent,
for $r$ being a few multiples of
$\xi_{eq}(t_{C}) $, for reasons that are not clear to us.  In that sense
Zurek's prediction is robust,
since explicit calculation\cite{ray2} shows that 
 $\xi (t)$ does not vary strongly in the interval
$-t_C \leq \Delta t\leq t_C$.

\subsection{QFT: the free roll}

The dynamical equations for a hot quantum field (Heisenberg
equations with Boltzmann boundary conditions) are very different
from the empirical TDLG equation above.
Fortunately, as for the condensed matter case, the interval
$-t_C \leq \Delta t\leq t_C$ occurs in the {\it linear} regime, 
when the {\it operator} equations
\begin{equation}
\frac{\partial^{2}\phi_{a}}{\partial t^{2}} = -\frac{\delta
F}{\delta\phi_{a}},
\label{op}
\end{equation}
for $F$ of Eq.\ref{FR2}, are solvable in terms of the mode functions
$\chi^{\pm}_{k}(t)$, satisfying  
\begin{equation}
\Biggl [ \frac{d^2}{dt^2} + {\bf k}^2 + m^2(t) \Biggr ]\chi^{\pm}_{k}(t)  =0,
\label{mode}
\end{equation}
subject to $\chi^{\pm}_{k}(t)
=
e^{\pm i\omega_{in}t}$ at
$t\leq 0$,
for incident frequency $\omega_{in} = \sqrt{{\bf
k}^{2} +\epsilon_{0} M^{2}}$, for $m^2(t) = \epsilon (t)M^{2}$,
where $\epsilon (t)$ is parametrised as for the  TDLG equation
above.  This corresponds to a temperature quench from an initial
state of thermal equilibrium at temperature $T_{0}>T_c $, where
$(T_{0}/T_{c} -1) = \epsilon_{0}$. There is no reason to take $\epsilon_{0}$ small. 
The diagonal correlation function $G(r, t)$ of Eq.\ref{diag} is given as the
equaltime propagator
\begin{equation}
G(r, t)=\int d \! \! \! / ^3 k 
\, e^{i {\bf k} . {\bf x} } \chi^{+}_{k}(t) \chi^{-}_{k}(t)C(k),
\label{modesum}
\end{equation}
where $ C(k) =\coth(\omega_{in} (k)/2T_0
)/2\omega_{in}(k)$ encodes the initial conditions.
 
An exact
solution can be given\cite{bowick} in terms of Airy, and related, functions. 
Dimensional analysis shows that, at displaced time $\Delta t =0$, 
$\xi_{eq}(t_{C})$ of Eq.\ref{xiC} again sets the scale of the equaltime
correlation function.  We have yet to work out whether the
coefficient is appproximately unity, but Kibble's insight is
correct.  Further detail is unnecessary since we shall argue that it
is the assumption that we can infer the defect density from
$\xi_{eq}(t_{C})$ that is flawed.

\section{Defect densities do not determine correlation lengths directly}

We have seen that there is no reason to disbelieve the causal
arguments of Kibble for QFT and Zurek for condensed matter as to the
time at which the field fluctuations freeze in, and the associated
correlation length.  The excellent agreement with the $^{3}He$
experiments also shows that, for condensed matter, this length does, indeed,
characterise vortex separation at the time when the defects form.

However, if we take the Lancaster experiment at face value, this
cannot always be the case. If we think about why this might be, the
differences between the $^{3}He$ and $^{4}He$ experiments are
twofold.  Firstly, the nuclear-driven quench rate for $^{3}He$ experiments is
several orders of magnitude faster than the mechanically-driven
quenches of the $^{4}He$ experiments.  Secondly, for $^{3}He$ the
Ginzburg regime is extremely narrow, whereas for $^{4}He$ it is very
broad.  In fact, the $^{4}He$ experiments begin and end in the Ginzburg
regime, where thermal fluctuations are important.  The causality
arguments are too simple to accomodate these facts.

If these differences are to be visible in the formalism, it can
only be through the way in which we relate  vortex density
to correlation length.  That is, the flaw of the prediction lies in the assumption
Eq.\ref{ndef}, as becomes apparent once we appreciate that it hides the problem of how to count
vortices.  One way is to work through line zeroes. For convenience
we restrict ourselves to $O(2)$ vortices whose cores are line zeroes of the complex
field $\phi$.  \footnote{More complicated vortices
require more complicated analysis, but we expect essentially the same conclusions.}
The converse is not true since zeroes occur on
all scales.  However, a starting-point for counting
vortices in superfluids
is to count
line zeroes of an appropriately coarse-grained field, in which
structure on a scale smaller than $\xi_{0}$, the classical vortex size, is
not present\cite{popov}.  
This is also the unstated
basis of the numerous numerical simulations\cite{tanmay} of cosmic
string 
networks built from Gaussian fluctuations (but see \cite{gleiser}). 
For the moment, we put in a cutoff $l = O(\xi_{0})$ by hand, as
\begin{equation}
G_{l}(r,\Delta t) = \int d \! \! \! / ^3 k\, e^{i{\bf k}.{\bf x}}G(k,\Delta t)\,e^{-k^{2}l^{2}}.
\label{Gl}
\end{equation}
We stress that  the {\it long-distance} correlation length
$\xi_{eq}(t_{C})$  depends essentially on the position of the
nearest singularity of $G(k,\Delta t)$ in the complex k-plane, {\it
independent} of $l$. 

This is not the case for the line-zero density $n_{zero}$, 
depending, in our Gaussian approximation\cite{halperin,maz}, on
the {\it short-distance} behaviour of $G_{l}(r,\Delta t)$ as
\begin{equation}
n_{zero}(\Delta t) = \frac{-1}{2\pi}\frac{G_{l}''(0,\Delta t)}{G_{l}(0,\Delta t )},
\label{ndef2}
\end{equation}
the ratio of fourth to second moments of $G(k,\Delta t)\,e^{-k^{2}l^{2}}$. 

\subsection{TDLG condensed matter}

In the linear regime everything is calculable.
If we define the line-zero separation $\xi_{zero}$ by
\begin{equation}
n_{zero}(\Delta t) = \frac{1}{2\pi\xi_{zero}(\Delta t)^{2}}
\label{nzdef}
\end{equation}
it is apparent that $\xi_{zero}(\Delta t)$ has little, if anything,
to do with $\xi_{eq}(\Delta t)$ directly.

In fact, at $\Delta t =0$,
\begin{equation}
\xi_{zero}^{2}\approx O(l^{2}),
\label{xidef0}
\end{equation} 
independent of $\epsilon_{0}$.  We have a situation in which the
density of line zeroes depends entirely on the scale at which we look. 
Such fractal behaviour cannot be
understood as representing stable vortices, although such line
zeroes might be termed {\it fluctuation} vortices. We would not wish to
identify the length scale $\xi_{zero}$ with
defect separation at this time, even if the latter could be defined.  

This is not surprising.  Although the field correlation length
$\xi (\Delta t)$ may
have frozen in by $\Delta t = 0$, the symmetry breaking has not been effected. 
In a classical sense at least, vortices can only be identified once
the field magnitude has grown to its equilibrium value
\begin{equation}
\langle |\phi |^{2}\rangle = \alpha_{0}/\beta,
\label{eq}
\end{equation} 
and we should not begin to count them before then.  Even though the
long-range correlation length $\xi (\Delta t)$ will not change
substantially in that time, this will not be the case for $\xi_{zero}(\Delta
t)$, as long-range modes increase in amplitude as the field becomes ordered.

Specifically, on continuing to use (\ref{tdlg}) for times $t >
\epsilon_{0}\tau_{Q}$ we see that,
as the unfreezing occurs, long wavelength modes with $k^{2} < t/\tau_Q -
\epsilon_{0}$ grow exponentially.  Provided
$\epsilon_{0}^{2}\tau_{Q}\gg 1$ they soon begin to dominate the
correlation functions.  Let 
\begin{equation}
G_{n}(\Delta t ) = \int_{0}^{\infty}dk\,k^{2n}\,G(k,\Delta t)
\end{equation}
be the moments of
$G(k,\Delta t)$.
In units of $\xi_{0}$ and $\tau_{0}$ we find\cite{ray} that
\begin{equation}
G_{n}(\Delta t)\approx\frac{I_{n}}{2^{n + 1/2}}\,e^{(\Delta t/{\bar t})^{2}}
\int_{0}^{\infty}dt'\,\frac{e^{-(t'-\Delta t)^{2}/{\bar t}^{2}}}{[t'
+l^{2}/2]^{n +1/2}},
\label{Gt}
\end{equation}
where we measure the dimensionless time $\Delta t$ in units of ${\bar t} = \sqrt{\tau_{Q}}$
from $t = \epsilon_{0}\tau_{Q}$ and $I_{n} = \int_{0}dk k^{2n}\, e^{-k^{2}}$.
For small relative times the integrand gets a large
contribution from the ultraviolet {\it cutoff dependent} lower endpoint, and we recover
(\ref{xidef0}).
As long as the endpoints make a significant contribution to the whole
then the density of line zeroes derived from (\ref{Gt}) will be
strongly dependent on scale.
Only when their contribution is small and $\partial n_{zero}/\partial l$ is
small in comparison to
$n_{zero}/l$ at $l = \xi_{0}$ can we identify the essentially
non-fractal $n_{zero}$
with a meaningful vortex density. 

$\Delta t_{1}$ and $\Delta t_{2}$,  the times at which the exponential modes begin
to dominate in the integrands of $G_{1}$ and $G_{2}$, can be determined 
by comparing the relative strengths of the contributions from the
scale-independent saddlepoint and the endpoint in (\ref{Gt}). 
If $p_{1}= \Delta t_{1}/{\bar t}$ and $p_{2}= \Delta t_{2}/{\bar t}$
are the multiples of ${\bar t}$ at which it happens then,
for the former to
dominate the latter requires 
$e^{p_{1}^{2}}/p_{1}^{3/2}> \tau_{Q}^{1/4}/\sqrt{2\pi}$ and
$e^{p_{2}^{2}}/p_{2}^{5/2}> 3\sqrt{2}\tau_{Q}^{3/4}/\sqrt{\pi}$.
We see that it takes longer for the long wavelength modes to
dominate $G_{2}$ than the order parameter $G_{1}$.
Because of the exponential growth, $p_{1}$ and $p_{2}$ are $O(1)$.
We stress that these are
lower bounds.

If the linear equation (\ref{Gt}) were valid for later times $\Delta
t>\Delta t_{2}$  then the integrals are dominated by the
saddle-point at $t'=\Delta t$, to give a separation of line
zeroes $\xi_{zero}(\Delta t )$ of the form
\begin{equation}
\xi_{zero}^{2}(\Delta t) = \frac{G_{1}(\Delta t)}{G_{2}(\Delta t)}
\approx\frac{4\Delta t}{3{\bar t}}\xi_{0}^{2}\bigg(\frac{\tau_{Q}}{\tau_{0}}\bigg)^{1/2}
=\frac{4\Delta t}{3{\bar t}}\xi_{eq}(t_{C})^{2}
\label{newZ}
\end{equation}
approximately {\it independent} of the cutoff $l$ for $l = O(\xi_{0})$.
Only then, because of the transfer of
power to long wavelengths, do  line zeroes become widely separated, and
$\xi_{zero}(\Delta t)$ does begin to measure vortices. 
Thus, if the order parameter is large enough that it takes a long
time (in units of ${\bar t}$) for the field to populate its
ground states (\ref{eq}) we would recover the result
$\xi_{zero}= O(\xi_{eq}(t_{C}))$ as an order of magnitude result from
(\ref{newZ}) even though, {\it a priori}, they are unrelated. 
This explains how experiments can be in agreement with the simple
causal predictions.

Whether we have time enough depends on the self-coupling $\beta$,
which determines when the linear approximation fails.
At the absolute latest, the correlation function
must stop its exponential growth at $\Delta t = \Delta t_{sp}$, when $\langle |\phi |^{2}\rangle$,
proportional to $G_{1}$, satisfies (\ref{eq}).  Let us suppose that
the effect of the backreaction that stops the growth initially
freezes in any defects.  This then is our prospective starting point for identifying
and counting vortices.  

It happens that (\ref{eq}) is satisfied when
\begin{equation}
G_{1} (\Delta t_{sp})= \pi^{2}\alpha_{0}^{2}\xi_{0}^{3}/\beta k_{B}T_{c}
= \pi^{2}/\sqrt{1-T^{-}_{G}/T_{c}},
\label{Gmax}
\end{equation}
where
$T^{-}_{G}$ is the Ginzburg temperature introduced earlier.  
 
In order that $G_{1}$ is dominated by the exponentially growing
modes, so as to recover the modification of prediction (\ref{ndef}) 
in the form (\ref{newZ}) for identifiable vortices, the condition $\Delta t_{sp}>\Delta t_{1},
\Delta t_{2}$ becomes, from (\ref{Gmax})
\begin{equation}
(\tau_{Q}/\tau_{0})(1-T^{-}_{G}/T_{c})<C\pi^{4},
\label{tsp}
\end{equation}
on restoring $\tau_{0}$, where $C\approx 1$.

How strongly the inequality should be satisfied in (\ref{tsp}) is not obvious,
assuming as it does that the backreaction is effectively
instantaneous, and that mean field critical indices are valid. 
However, there is no way that the inequality can be remotely
satisfied for $^{4}He$, when subjected to a slow mechanical quench, as in
the Lancaster experiment, for which $\tau_{Q}/\tau_{0} =
O(10^{10})$, since the Ginzburg regime is so large 
that $(1-T_G /T_c ) = O(1)$.  
That is, field growth must stop long before  vortices are  well
defined.  Further, since the experiment leaves the superfluid in the
Ginzburg regime, thermal fluctuations will inhibit the creation of stable defects,
although it may be that
incoherent fluctuations, even if not vortices, will give a signal.   
Adopting renormalisation improved critical indices cannot repair
such a deficit.
Whatever, there is no
reason to expect a vortex density (\ref{ndef2}).

The situation for $^{3}He-B$ is potentially very different,  
For rapid quenches a 
 TDLG approach is valid\cite{Bunkov}. 
Suppose that (\ref{tsp}), which looks like an inequality between
the quench rate and the equilibriation rate that permits unstable
modes time to dominate over short-range fluctuations, remains true, albeit
with new coefficients.
Firstly, for $^{3}He-B$, 
the Ginzburg regime is very small, with $1-T_{G}/T_{c} =
O(10^{-8})$.  The quench ends outside it.
Secondly, in generating the phase transition by nuclear reactions,
rather than by mechanical expansion, 
the quench rate is increased dramatically, with 
$\tau_{Q} = 10^{2} - 10^{4}$.
The inequality (\ref{tsp}) is satisfied by a huge margin and we can understand the success of the Helsinki and
Grenoble experiments.

\subsection{QFT: mode growth v fluctuations}

It is because the formation of defects is an early-time occurrence that
it is, in large part, amenable to analytic solution.  Again we
revert to the mode decomposition of Eq.\ref{mode}.  The field
becomes ordered, as before, because of the exponential growth of
long-wavelength modes, which stop growing once the field has sampled
the groundstates.  What matters is the relative weight of these
modes (the 'Bragg' peak)
to the fluctuating short wavelength modes in the decomposition
Eq.\ref{modesum} at this time, since the contribution of these latter is very
sensitive to the cutoff $l$.  Only if their contribution to
Eq.\ref{ndef} is small when field growth stops can a network of vortices be well-defined at
early times, let alone have the predicted density.  
Since the peak is non-perturbatively large this
requires small coupling, which we assume.  

However, to determine the counterpart to Eq.\ref{tsp} for QFT is
not so simple, and it helps to begin with an idealised problem, in
which the quench is {\it instantaneous}
($t_Q =0$), and $m^{2}(t)$ flips from $+\epsilon_{0}M^{2}$ to $-M^{2}$ at $t=0$.
Eq.\ref{ndef} is clearly inappropriate,
a useful warning that the simple picture above has its limitations.   
However, once we have understood the instantaneous case the slow
quench is not much more difficult.

The mode equations Eq.\ref{mode} are instantly solvable, and $G(r,
t)$ of Eq.\ref{modesum} is simple calculable. 

Rather than introduce a cutoff at scale $l$, $l^{-1} = \Lambda =
O(M)$, as in Eq.\ref{Gl}, we adopt the more brutal approach of
cutting off modes at $|{\bf k}|<\Lambda $ in Eq.\ref{modesum}.
Imposing the KMS  boundary conditions at 
$t\leq 0$ for an initial Boltzmann
distribution at temperature $T_0$ compatible with thermal mass $\sqrt{\epsilon_{0}}M$
 determines  the resulting coarse-grained 
$G_{\Lambda}(r;t)$ uniquely. For $t<0$ all modes are oscillatory,
but for $t>0$ long wavelength modes with $k^{2}>\Lambda^{2}$ are
unstable, and grow exponentially.  For $\Lambda=O(M)$, but $\Lambda >M$,
say, $G_{\Lambda}(r;t)$ can be decomposed in an obvious way as
$G_{\Lambda}(r;t) = G^{in}_{\Lambda}(r)$ for $t\leq 0$, and
\begin{equation}
G_{\Lambda}(r;t) = G^{in}_{\Lambda}(r) + G_{|{\bf k}|<M}(r;t) + G_{\Lambda >|{\bf k}|>M}(r;t)
\label{Wdecomp2}
\end{equation}
for $t>0$, where\cite{boyanovsky} the second and third terms are the long and
short wavelength fluctuations that grow from $t=0$ onwards.

The equilibrium background term $G_{\Lambda}^{in}(r)$ has the form
\begin{equation}
G_{\Lambda}^{in}(r)=\int_{|{\bf k}|<\Lambda } d \! \! \! / ^3 k 
\, e^{i {\bf k} . {\bf x} } C(k),
\end{equation}
where $ C(k)$ encodes the initial conditions as before.
The correlation length of the field is $\xi_{0} = M^{-1}$ and, for cutoff
$\Lambda = O(M)$, the cold vortex thickness, we have line zero density
$n_{zero} =  O(\Lambda^2 )$, as in the condensed matter case.  
There is a sea of line zeroes,
largely tiny loops, 
separated only by the
Compton wavelength.  The oscillating field fluctuations make them
unsuitable  as serious candidates for string since the density depends
critically on the scale $\Lambda^{-1}$ at which we view the field.  

For early positive times $n_{zero}$ is equally sensitive to
$\Lambda$ but, 
once $Mt\gg 1$, the relevant term is $G_{|{\bf k}|<M}(r;t)$,
whose integral at time t is dominated by the nonperturbatively large
peak\cite{boyanovsky} in the power of the fluctuations at $k$ around
$k_0$, where
$t k_0^2\simeq M$.  Once $k_{0}\ll M $ we have the required
insensitivity of the line density to the scale at which we coarsegrain.
We find, approximately, that
\begin{equation}
G_{|{\bf k}|<M}(r;t)\propto\frac{MT_0}{(tM)^{3/2}}e^{2Mt}\exp\{-r^{2}M/4t\}.
\label{Wapp}
\end{equation}
The prefactor comes from approximating $C(k_{0})$ by  $T_0/\epsilon_{0}M^{2}$.
For $t>0$ the
unstable modes grow until $G_{|{\bf k}|<M}(0;t_{sp}) =O(\phi^{2}_{0}) = O(M^{2}/\lambda )$,
determining
the spinodal time $t_{sp}$  as   $Mt_{sp}=O(\ln (1/\lambda ))$.
The long-distance behaviour of $G^{in}_{\Lambda}(r) =
O(e^{-Mr})$, with its correlation length $\xi = M^{-1}$ is a
shortlived relic of the initial thermal conditions.  After a time $t_r
= O(M^{-1})\leq t_{sp}$, it is rapidly
supplanted by the behaviour of the expanding long wavelength modes 
$G_{|{\bf k}|<M}(r;t)/G_{|{\bf k}|<M}(0;t)\approx\exp\{-r^{2}/\xi^{2}(t)\}$ where
$\xi^{2}(t) = 4t/M\approx 4/k_{0}^{2}$.
With  $G_{|{\bf k}|<M}(0;t_{sp}) = O(\lambda^{-1})$ non-perturbatively
large and
$G^{in}_{\Lambda}(0)$ and $ G_{\Lambda >|{\bf k}|>M}(0;t_{sp})$ of
order $\lambda^{-1/2}$,
\begin{equation}
n_{zero}(t_{sp})\approx  
\frac{1}{\pi\xi_{0}^{2}\ln (1/\lambda)}[1 + O(\lambda^{1/2}\ln (1/\lambda ) )].
\label{niMf}
\end{equation}
The relative error term in the brackets is, in large part, a measure of the fluctuation
vortices that we mentioned earlier and is a measure of the stability of
this density to changes in the
coarse-graining scale $\Lambda = O(M)$.  For weak coupling
$O(\lambda^{1/2}\ln
(1/\lambda ))\ll 1$.

There are other, less direct, ways to understand why the strings only become
well-defined once $G_{M}(0;t)$ is nonperturbatively large, even
though the density is
independent of its {\it magnitude}. Instead
of a field basis, we can work in a particle basis and measure the
particle production as the transition proceeds.  
Whether we expand with respect to the original
Fock vacuum or with respect to the adiabatic vacuum state, the
presence of a non-perturbatively large peak in $k^2 G(k;t)$ signals
a non-perturbatively large occupation number $N_{k_{0}}\propto 1/\lambda$
of particles at the same wavenumber $k_0$\cite{boyanovsky}. 
With $n_{zero}$ of (\ref{niMf}) of order $k_{0}^{2}$ 
this shows that the long wavelength modes can now begin to be treated classically.  

From a slightly different viewpoint, the Wigner functional only
peaks about the classical phase-space trajectory once the power is
non-perturbatively large\cite{guth,muller} from time $t_{sp}$ onwards.  
More crudely, the diagonal density matrix
elements (field probabilities) are only then significantly non-zero for
non-perturbatively large field configurations
$\phi\propto\lambda^{-1/2}$ like vortices.

So far we have done no more than estimate the density of
coarse-grained line zeroes for a free roll at the time $t_{sp}$ at which
 nothing has frozen in.  
For this {\it global} $O(2)$ theory the damping of domain growth occurs
by the self-interaction effectively  forcing the
negative $m^{2}(t)$ to
vanish so as to produce Goldstone particles.  This initial slowdown
leads to no qualitative change
in the vortex density.

All the results above were for the instantaneous quench. We now
return to the original problem of slower quenches, with $\epsilon
(t)$ as in Eq.\ref{eps},
in which the symmetry-breaking begins at relative time $\Delta t = 0$..
For a {\it
free} roll,  
the exponentially growing modes that appear when
$\Delta t>t^{-}_{k} = t_{Q}k^{2}/M^{2}$ lead to the approximate solution\cite{ray2}
\begin{equation}
G(r;\Delta t)\propto
\frac{T}{M|m(\Delta t)|}\bigg(\frac{M}{\sqrt{\Delta
tt_Q}}\bigg)^{3/2}e^{\frac{4M\Delta t^{3/2}}
{3\sqrt{t_Q}}}
e^{-r^{2}/\xi^{2}(\Delta t)}
\label{Wexp}
\end{equation}
where 
$\xi^{2}(\Delta t) = 2\sqrt{\Delta tt_Q}/M$.
The provisional relative freeze-in time $\Delta t_{sp}$ is then,  for 
$Mt_{Q} < (1/\lambda)$,
\begin{equation}
M\Delta t_{sp} \simeq (Mt_{Q})^{1/3}(\ln (1/\lambda ))^{2/3}
\simeq Mt_{C}(\ln (1/\lambda ))^{2/3}.
\label{tfs}
\end{equation}
This is greater than $Mt_{C}$, but not by a large multiple unless we
have a superweak theory.  More detail is given in \cite{ray2}.

At this qualitative level the correlation length at the spinodal time is
\begin{equation}
M^{2}\xi^{2}(t_{sp})\simeq (Mt_{Q})^{2/3}(\ln (1/\lambda ))^{1/3}.
\label{chiss2}
\end{equation}
The effect of the other modes is larger than for the instantaneous
quench,  giving, at $t=t_{sp}$
\begin{equation}
n_{zero}=  \frac{M^{2}}{\pi (M\tau_{Q})^{2/3}}
(\ln (1/\lambda ))^{-1/3}[1 + E].
\label{nisMf}
\end{equation}
The error term $E =O(\lambda^{1/2}(Mt_{Q})^{4/3}(\ln
(1/\lambda))^{-1/3})$ is due to fluctuation vortices.  
In mimicry of Eq.\ref{ndef} it is helpful to rewrite
Eq.\ref{nisMf} as
\begin{equation}
n_{zero}=  \bigg[\frac{1}{\pi \xi_{0}^{2}}\bigg(\frac{\tau_{0}}{\tau_{Q}}\bigg)^{2/3}\bigg]
(\ln (1/\lambda ))^{-1/3}[1 + E].
\label{nisMf2}
\end{equation}
in terms of the scales $\tau_{0} = \xi_{0} = M^{-1}$.
The first term in Eq.\ref{nisMf2} is the Kibble estimate of
Eq.\ref{ndef}, the second is the small multiplying factor, rather like
that in Eq.\ref{newZ}, that yet again shows that estimate can be
correct, but for completely different reasons.  The third term shows
when it can be correct, since $E$ is also a measure of the
sensitivity of $n_{zero}$ to the scale at which it is measured.
We note that if we take $M\tau_{Q} = \ln (1/\lambda )$ we recover the
instantaneous results qualitatively.  Only for larger $\tau_{Q}$
will the quench give different results.  The condition $E^{2}\ll 1$,
necessary for a vortex network to be defined, is then guaranteed if
\begin{equation}
(\tau_{Q}/\tau_{0})^{2}(1-T^{-}_{G}/T_{c})<1,
\end{equation}
on using the relation $(1-T^{-}_{G}/T_{c}) = O(\lambda )$.  This is
the QFT counterpart to Eq.\ref{tsp}.

The density of  Eq.\ref{nisMf2} is {\it smaller} than that of
Eq.\ref{nisMf} by the factor $(\ln (1/\lambda ))^{1/3}$ which 
 is, in principle, a large number even if in
practice it is only a small multiple.
The easiest way to enforce $E\ll 1$ and  $M\tau_{Q} > \ln (1/\lambda
)$ is to take $M\tau_{Q} = \ln (1/\lambda )^{\alpha}$, for $\alpha
>1$. The effect in Eq.\ref{nisMf2} is merely to renormalise the
critical index.  Of course, the Kibble prediction Eq.\ref{ndef} was
only an estimate.  Although it is good qualitatively it
 is  misleading  when considering definition of the
network since a simple calculations shows that, at time $t_C$,
the string density is still XStotally sensitive to the definition of coarse-graining.

Finally, suppose that this approach is relevant to the local strings
of a strong Type-II $U(1)$ theory for the early universe, in which
the time-temperature relationship $tT^{2} = \Gamma M_{pl}$ is valid,
where we take $\Gamma = O(10^{-1})$ in the GUT era.  If $G$ is
Newton's constant and $\mu$ the classical string tension then,
following \cite{zurek1}, $Mt_{Q}\sim
10^{-1}\lambda^{1/2}(G\mu)^{-1/2}$.  The dimensionless quantity 
$G\mu\sim 10^{-6} - 10^{-7}$ is
the small parameter of cosmic string theory.  A value $\lambda\sim
10^{-2}$ gives $Mt_{Q}\sim (Mt_{sp})^{a}, a\sim 2$, once factors of
$\pi$, etc.are taken into account, rather than
$Mt_{Q}\sim 1/\lambda$, and the density of Eq.\ref{nisMf2} may be relevant.

We have not attempted to show what happens directly after the formation of
defects for those cases  when they are well-defined by the time the
initial population of the ground states has occurred.  We have only
attempted to test the Kibble/Zurek scenarios for defect formation. For
strongly-damped systems like condensed matter systems the linear
approximation seems to be valid for longer times than might have
been guessed.  That would explain why the final population is simply
related to the initial population. For QFT it is much less obvious that
this is the case.  That is,
even when Kibble is correct, the later evolution may make the
estimate irrelevant.  To take one extreme example.  If we suppose
that the field evolution can be approximated by {\it self-consistent}
linearisation then, very rapidly, the scattering between field modes,
absent in the linear approximation here, will be enough to
redistribute power back to short wavelengths.  The density of line
zeroes then begins to {\it increase} rapidly.  If they do represent
strings their density is likely to be independent of $\tau_{Q}$\cite{devega}.  

\section{Conclusions}

The Kibble/Zurek scenario for the onset of a continuous transition,
either in QFT or condensed matter,
suggests that causality determines the correlation length of the
fields at the time that they freeze in.  Our simple calculations show
that this is qualitatively correct.

A further step in their arguments was to identify this calculable correlation
length with the length of defect separation (cosmic strings or
vortices) at the time at which such defects were produced.  Whereas
this is in agreement with vortex experiments on $^{3}He$ it is not the case
for $^{4}He$.  
In order to understand this we have identified defect density with
field-zero density, once this is effectively independent of the
scale at which the zeroes are counted.  

The separation of zeroes
is determined from the short-distance field correlations and the
correlation length (by definition) from the long-distance field correlation
function.  Nonetheless, the field correlation length at the time it
freezes in and the zero separation length when the initial relaxation of the
fields to their ground states is complete are qualitatively the same, {\it
provided} thermal fluctuations are not too strong for a well-defined 
vortex network to be established.
Qualitatively, this requires (in the terminology of the text)
\begin{equation}
(\tau_{Q}/\tau_{0})^{\gamma}(1-T^{-}_{G}/T_{c})<1,
\label{end}
\end{equation}
where (in mean field) $\gamma = 1$ for condensed matter, and $\gamma
=2$ for QFT.

This linking of quench rate to the Ginzburg regime explains why $^{3}He$
and $^{4}He$ experiments are so different, and why we expect the
simple predictions for vortex densities not to be verified in the
latter. A constraint like Eq.\ref{end} is not strong enough to
preclude initial defect formation in the early universe.

\section*{Acknowledgements}

I particularly thank Glykeria Karra, with whom most of this work
was done.  I also would like to thank Alasdair Gill, Tom Kibble, 
Wojciech Zurek and Yuriy Bunkov for fruitful discussions.
This work is the result of a network supported by the European
Science Foundation.


\begin{thebibliography}{99}

\bibitem{kibble1} T.W.B. Kibble, {\it J. Phys.} {\bf A9}, 1387 (1976).

\bibitem{kibble2} T.W.B. Kibble, in {\it Common Trends in Particle and
Condensed Matter Physics},  {\it Physics Reports} {\bf 67}, 183 (1980).

\bibitem{zurek1} W.H. Zurek, {\it Nature} {\bf 317}, 505 (1985), 
{\it Acta Physica Polonica} {\bf B24}, 1301 (1993).
See also W.H. Zurek, {\it
Physics Reports} {\bf 276}, Number 4, November 1996 

\bibitem{zurek2} W.H. Zurek, {\it Nature} {\bf 382}, 297 (1996),\\
P. Laguna and W.H. Zurek, {\it Phys. Rev. Lett.} {\bf 78}, 2519 (1997).
A. Yates and W.H. Zurek, hep-ph/9801223.

\bibitem{volovik2} M.M. Salomaa and G.E. Volovik, {\it Rev. Mod. Phys.}
{\bf 59}, 533 (1987)


\bibitem{helsinki} V.M.H. Ruutu {\it et al.}, {\it Nature} {\bf 382}, 334 (1996).

\bibitem{grenoble} C. Bauerle {\it et al.}, {\it Nature} {\bf 382},
332 (1996).

\bibitem{lancaster} P.C. Hendry {\it et al}, {\it Nature} {\bf 368}, 315 (1994).

\bibitem{lancaster2} P.V.E McClintoch {\it et al.}, to be published
in {\it Phys. Rev. Letters}.

\bibitem{ray}G. Karra and R. J. Rivers, to be published in {\it Phys.
Rev. Letters}.

\bibitem{bowick}M. Bowick and A. Momen, preprint hep-ph/9803284, to
be published.

\bibitem{popov} V.N. Popov {\it Functional integrals and collective
excitations}, Cambridge University Press (1997), Chapter 8. 

\bibitem{tanmay} T. Vachasparti and A. Vilenkin, {\it Phys. Rev.}
{\bf D30}, 2036 (1984), A.Vilenkin and R.J. Scherrer, {\it Phys. Rev.} {\bf D5}
647 (1997), see also preprint hep-ph/9709498.

\bibitem{gleiser} M. Gleiser and H-R. Muller, preprint
hep-lat/9704005, to be published in {\it Physics Letters B}.


\bibitem{halperin} B.I. Halperin, published in {\it Physics of Defects},
proceedings of Les Houches, Session XXXV 1980 NATO
ASI, editors Balian, Kl\'{e}man and Poirier 
(North-Holland Press, 1981) p.816.

\bibitem{maz} F. Liu and G.F. Mazenko, {\it Phys. Rev.} {\bf B46}, 5963 (1992).

\bibitem{Bunkov} Yu. M. Bunkov and O.D. Timofeevskaya, preprint
cond-mat/9706004, {\it Phys.
Rev. Letters}, to be published.

\bibitem{boyanovsky} D. Boyanovsky, Da-Shin Lee and A. Singh, 
{\it Phys. Rev.} {\bf D48}, 800 (1993), \\
D. Boyanovsky, H.J. de Vega and R. Holman, {\it Phys.
Rev.} {\bf D49}, 2769 (1994).


\bibitem{guth} A. Guth and S-Y. Pi, {\it Phys. Rev.} {\bf D32},
1899 (1985).  O. Eboli, R. Jackiw and S-Y. Pi, {\it Phys. Rev.} {\bf
D37}, 3557 (1988).


\bibitem{muller} S. Mrowczynski and B. Muller, {\it Phys. Rev.} {\bf
D50}, 7542 (1994).

\bibitem{ray2} G. Karra and R.J. Rivers, {\it Physics Letters} {\bf B414},
28 (1997)

\bibitem{devega}H.J. de Vega et al., private communication.

\end{thebibliography}
\end{document}